\documentclass[11pt]{article}
\usepackage{latexsym}  
\usepackage{amssymb}
\usepackage{graphicx}
\usepackage{amsmath}

\begin{document}

\begin{center}

{\Large\bf Origin of Hierarchical Structures of}

 {\Large\bf  Quark and Lepton Mass Matrices}

\vspace{4mm}

{\bf Yoshio Koide$^a$ and Hiroyuki Nishiura$^b$}

${}^a$ {\it Department of Physics, Osaka University, 
Toyonaka, Osaka 560-0043, Japan} \\
{\it E-mail address: koide@kuno-g.phys.sci.osaka-u.ac.jp}

${}^b$ {\it Faculty of Information Science and Technology, 
Osaka Institute of Technology, 
Hirakata, Osaka 573-0196, Japan}\\
{\it E-mail address: hiroyuki.nishiura@oit.ac.jp}

\date{\today}
\end{center}

\vspace{3mm}

\begin{abstract}
It is shown that the so-called ``Yukawaon" model 
can give a unified description of masses, mixing and $CP$ violation parameters of 
quarks and leptons without using any hierarchical (family number-dependent) 
parameters besides the charged lepton masses. 
Here, we have introduced  a phase matrix $P={\rm daig}(e^{i \phi_1}, e^{i \phi_2}, 
e^{i \phi_3})$ with the phase parameters 
$(\phi_1, \phi_2, \phi_3)$ which are described in terms of  family number-independent parameters,  
together with using only the charged lepton mass parameters 
as the family number-dependent parameters.  
  In this paper, the  $CP$ violating phase 
parameters $\delta_{CP}^q$ and $\delta_{CP}^\ell$ in the standard 
expression of $V_{CKM}$ and $U_{PMNS}$ are predicted as $\delta_{CP}^q \simeq 72^\circ$ 
and $\delta_{CP}^\ell \simeq -76^\circ$, respectively, 
i.e. $\delta_{CP}^\ell \sim  - \delta_{CP}^q$.  

\end{abstract}

PCAC numbers:  
  11.30.Hv, 
  12.15.Ff, 
  14.60.Pq,  
  12.60.-i, 

\vspace{3mm}

\noindent{\large\bf 1 \ Introduction}

\vspace{2mm}

{\bf 1.1 \ What is the Yukawaon model}

One of the big subjects in the particle physics is to investigate 
the origin of flavors.
There is an attractive idea that the flavor physics is understood 
from the point of view of a family symmetry \cite{f-symmetry}.  
However,  the symmetry has to be explicitly broken by the Yukawa 
coupling constants $Y_f$ ($f= \nu, e, u, d$) 
if we suppose the family symmetry to be a continuous symmetry.  
Therefore, the symmetry is usually considered as a discrete symmetry. 
If we adhere to the basic idea that the flavor symmetry should be 
a continuous symmetry which is unbroken at the start, we are forced 
to consider that the Yukawa coupling constants are effective
coupling constants $Y_f^{eff}$ which are given by vacuum expectation 
values (VEVs) of scalars (``Yukawaons")  $Y_f$ with $3 \times 3$ components
\cite{YK_09}:
$$
(Y_f^{eff})_i^{\ j} = \frac{y_f}{\Lambda} \langle Y_f\rangle_i^{\ j}  
\ \ \ \ (f=u, d, \nu, e),
\eqno(1.1)
$$
where $\Lambda$ is an energy scale of the effective theory. 
In the Yukawaon model, all the flavons \cite{flavon} are expressed 
by $3 \times 3$ components of 
U(3). We consider no substructures of it such as $2 \times 2$ and so on.
 
In the Yukawaon model, we assume a U(3) family symmetry, 
and take the following would-be Yukawa interactions:
$$
H_Y = \frac{y_\nu}{\Lambda} (\bar{\ell}_L)^i (\hat{Y}_\nu)_i^{\ j} (\nu_R)_j  H_u 
+ \frac{y_e}{\Lambda} (\bar{\ell}_L)^i (\hat{Y}_e)_i^{\ j} (e_R)_j  H_d 
+ y_R (\bar{\nu}_R)^i (Y_R)_{ij} (\nu_R^c)^j 
$$
$$
+ \frac{y_u}{\Lambda}  (\bar{q}_L)^i (\hat{Y}_u)_i^{\ j} (u_R)_j H_u 
+ \frac{y_d}{\Lambda}  (\bar{q}_L)^i (\hat{Y}_d)_i^{\ j}  (d_R)_j H_d  ,
\eqno(1.2)
$$
where $\ell_L=(\nu_L, e_L)$ and $q_L=(u_L, d_L)$ are SU(2)$_L$ doublets. 
$H_u$ and $H_d$ are two Higgs doublets. 
The third term in Eq.(1.2) leads to the so-called neutrino seesaw mass 
matrix \cite{seesaw} 
$M_\nu =\hat{Y}_\nu Y_R^{-1} \hat{Y}^T_\nu$, where $\hat{Y}_\nu$ and $Y_R$ 
correspond to the Dirac and Majorana mass matrices of neutrinos, respectively.  
Hereafter, for convenience, we use notation 
$\hat{A}$, $A$ and $\bar{A}$ for fields with ${\bf 8}+{\bf 1}$,
${\bf 6}$ and ${\bf 6}^*$ of U(3), respectively.

In order to distinguish each Yukawaon from the others, we assume that
$\hat{Y}_f$ have different $R$ charges from each other by considering 
$R$-charge conservation [a global U(1) symmetry in $N=1$ supersymmetry (SUSY)]. 
Of course, the $R$-charge conservation is broken
at an energy scale $\Lambda$, at which the U(3) family symmetry 
is broken.
For $R$ parity assignments, we inherit those 
in the standard SUSY model, i.e.  
$R$ parities of yukawaons $\hat{Y}_f$ (and all flavons) are the same 
as those of Higgs particles 
(i.e. $P_R({\rm fermion})=-1$ and $P_R({\rm scalar})=+1$), 
while quarks and leptons are assigned to 
$P_R({\rm fermion})=+1$ and $P_R({\rm scalar})=-1$.

A remarkable characteristic of the Yukawaon model is that it is possible 
to understand the observed hierarchical structures 
of masses and mixings of quarks and leptons without using 
any family number-dependent parameters except for the charged lepton 
masses. That is, all the quark and lepton masses and mixings can be 
understood in terms of only the observed charged lepton masses.   
In the Yukawaon model so far, our aim seems to be almost accomplished 
except for the following problem only: Namely, 
we have obliged to introduce a phase matrix 
$\langle P \rangle = {\rm diag} (e^{i\phi_1},  e^{i\phi_2}, e^{i\phi_3})$ 
in order to give a good fitting for the Cabibbo-Kobayashi-Maskawa (CKM) 
mixing matrix \cite{CKM}, where $(\phi_1, \phi_2, \phi_3)$ have been 
introduced as family number-dependent parameters.
In this paper, however, we will relate those parameters 
$(\phi_1, \phi_2, \phi_3)$ to the observed charged lepton masses 
$m_{ei}$ as discussed in Sec.4. 

Relations among Yukawaon VEVs $\langle\hat{Y}_f \rangle$ are obtained 
by supersymmetric vacuum conditions from U(3) symmetric and 
$R$-charge conserved superpotential.
In this article, the Yukawaon VEVs 
$\langle\hat{Y}_f \rangle$ are related to VEVs of fundamental flavons
$\Phi_f$ with a common bilinear form to all flavors:\footnote{
In earlier Yukawaon models \cite{Yukawaon_1}, 
the bilinear form was only for the up-quark sector, 
and the model could excellently lead to the so-called 
tribimaximal mixing \cite{tribi} in the lepton mixing 
under the use of only a few parameters. 
However, the model \cite{Yukawaon_1} (and also  \cite{Yukawaon_2}) could 
not give the observed sizable mixing angle $\theta_{13}$.  
We found that we can give the observed value
$\sin^2 2\theta_{13} \simeq 0.1$  \cite{theta13} only when we consider that all 
VEV relations are given by a common bilinear form Eq.(1.3) \cite{Yukawaon_3}. 
}   
$$
\langle \hat{Y}_f \rangle = k_f \langle \Phi_f \rangle 
\langle \bar{\Phi}_f \rangle + \xi_f {\bf 1} \ \ \ \ (f=u, d, \nu, e),
\eqno(1.3)
$$
where ${\bf 1}={\rm diag}(1,1,1)$.  
Here,  the VEV matrices $\langle \Phi_f \rangle$ and 
$\langle \bar{\Phi}_f \rangle$ are commonly related to 
a fundamental flavon VEVs $\langle \Phi_0 \rangle$  and 
$\langle \bar{\Phi}_0 \rangle$ by
$$
\begin{array}{l}
\langle \bar{P}_f \rangle^{ik}  \langle \Phi_f\rangle_{kl}  
\langle \bar{P}_f\rangle^{lj} = k'_f 
 \langle \bar{\Phi}_0 \rangle^{i\alpha}  \langle S_f \rangle_{\alpha\beta} 
\langle \bar{\Phi}_0 \rangle^{\beta j} , \\
\langle {P}_f\rangle_{ik}  \langle \bar{\Phi}_f\rangle^{kl} 
\langle {P}_f\rangle_{lj} 
= k'_f 
 \langle {\Phi}_0\rangle_{i\alpha}  \langle \bar{S}_f \rangle^{\alpha\beta}
 \langle {\Phi}_0\rangle_{\beta j}  , 
\end{array} \ \ \ \ (f=u, d, \nu, e).
\eqno(1.4)
$$
where $i$ and $\alpha$ are indices of U(3) and
U(3)$'$, respectively.  
For the VEV structures of $P_f$ and $\bar{P}_f$ in Eq.(1.4) and the $\xi_f$ terms 
in Eq.(1.3), we discuss in the next section.

The VEV structures $\langle S_f \rangle$ and $ \langle \bar{S}_f \rangle$ 
in Eq.(1.4) are given by 
\footnote{
The form (1.5) was suggested  by a ``democratic universal seesaw" 
mass matrix model \cite{K-F_ZPC96}, 
in which quark mass matrices are given by a form 
$\langle \Phi_e \rangle ( {\bf 1} + a_f X_3 )\langle \Phi_e \rangle$.
}
$$
\langle S_f \rangle = \langle \bar{S}_f \rangle = {\bf 1} + a_f X_3 ,
\eqno(1.5)
$$
where
$$
{\bf 1} = \left( 
\begin{array}{ccc}
1 & 0 & 0 \\
0 & 1 & 0 \\
0 & 0 & 1 
\end{array} \right) , \ \ \ \ \ 
X_3 = \frac{1}{3} \left( 
\begin{array}{ccc}
1 & 1 & 1 \\
1 & 1 & 1 \\
1 & 1 & 1 
\end{array} \right) . 
\eqno(1.6)
$$
The form of Eq.(1.6) is understood by a symmetry breaking
U(3)$' \rightarrow $S$_3$.


\vspace{2mm}

{\bf 1.2 Charged lepton sector as a fundamental flavor basis}

We consider that the charged lepton mass matrix is the most fundamental 
one compared with other mass matrices and that the charged lepton mass 
values play an essential 
role in understanding the flavor physics. The points of our postulation are as follows:

\noindent
(i) There is a fundamental flavon $\Phi_0$, and 
the reference basis in the flavor physics is defined by the diagonal 
basis of $\langle \Phi_0 \rangle$ and $\langle \bar{\Phi}_0 \rangle$:
$$
\langle {\Phi}_0 \, \rangle = \langle \bar{\Phi}_0 \rangle 
 \equiv v_0 \, {\rm diag}(x_1, x_2, x_3). 
 \eqno(1.7)
$$
where $x_i$ are real parameters with $x_1^2+x_2^2+x_3^3=1$.

\noindent
(ii) In the reference basis, the U(3)$'$ family symmetry is broken into S$_3$, 
i.e. VEVs of flavons $S_f$ and $\bar{S}_f$ take the form (1.5).

\noindent
(iii) The charged lepton mass matrix $\langle \hat{Y}_e \rangle$
should be diagonal and real as well as $\langle \Phi_0 \rangle$ 
and $\langle \bar{\Phi}_0 \rangle$, 
and it should be described only in terms of the fundamental 
parameters $x_i$.
Therefore, with demanding simplicity too, we require 
$$
a_e=0, \ \ \ \ \xi_e = 0.
\eqno(1.8)
$$
This means $x_i \propto m_{ei}^{1/4}$ 
($m_{ei}=(m_e, m_\mu, m_\tau)$).
In Sec.3, we use the following parameter values of $x_i$, 
$$
(x_1, x_2, x_3) =(0.115144, 0.438873, 0.891141) .
\eqno(1.9)
$$
In (1.9), we have used running mass values 
$m_e(\mu)=0.000486847$ GeV, $m_\mu(\mu)=0.102751$ GeV and 
$m_\tau(\mu) =1.7467$ GeV as the charged lepton mass values at 
$\mu=M_Z$, because our numerical predictions 
in the quark mass ratios are done at $\mu= M_Z$.
Note that the mass values $(m_e, m_\mu, m_\tau)$ have a large 
hierarchical structure, i.e. $m_e/m_\tau \sim 10^{-4}$, while
the values (1.9) have a mild hierarchical structure, i.e. 
$x_1/x_3 \sim 10^{-1}$.

In this paper, we do not ask the origin of the value $(x_1, x_2, x_3)$.
However, for reference, in Appendices A and B, we will demonstrate 
an example of the charged lepton mass relation in the present 
Yukawaon model.

\vspace{2mm}

{\bf 1.3 What is new}

The new characteristic points of the present Yukawaon model are as follows:

\noindent (i)
The bilinear form of the Yukawaon VEVs  given in Eqs.(1.3) and (1.4) has been 
adopted for some flavor sectors in 
the previous paper \cite{Yukawaon_3}, too.
However, in the present paper, we apply it to all the sectors $f=u, d, \nu, e$.

\noindent (ii)
So far, we have needed a VEV matrix $P$ given by
$$
P={\rm diag}(e^{i\phi_1}, e^{i\phi_2}, e^{i\phi_3}),
\eqno(1.10)
$$ 
in order to fit CKM mixing matrix $V_{CKM}$ reasonably.  
However, our aim was to describe all the masses and mixing of quarks and leptons
in terms of family number-independent parameters expect for the charged 
lepton masses. Therefore, 
the phase parameters $\phi_i$ were  against our aim and unwelcome as it is.
In the present paper, we try to denote the family number-dependent parameters 
$\phi_i$  in terms of  the observed charged lepton masses $m_{ei}$.
The details are discussed in Sec.4.

\noindent (iii) 
In general, the phase matrix $P$ affects not only 
the CKM quark mixing matrix  $V_{CKM}$
but also the Pontecorvo-Maki-Nakagawa-Sakata \cite{PMNS} 
(PMNS) lepton mixing matrix $U_{PMNS}$. 
Furthermore, predicted value of the $CP$ violating phase parameter 
$\delta_{CP}^\ell$ in $U_{PMNS}$ will depend on an appearing position of  
$P$ which is arbitrary at present. 
In the present paper, as we will discuss in Sec.2, we construct a model 
in which the phase matrix $P$ appears in the up-quark sector only, and 
it affects not only the  $V_{CKM}$ but also the $U_{PMNS}$ through 
Eq.(2.5) as we give later. 
As a result, as we discuss in Sec.3, we predict $CP$ violation 
phase parameters $\delta_{CP}^q$ and $\delta_{CP}^\ell$
in the standard expression of $V_{CKM}$ and $U_{PMNS}$ 
as $\delta_{CP}^\ell \simeq - \delta_{CP}^q \simeq -70^\circ$ unlike 
the previous papers.

In Sec.2, we construct a mass matrix model base on a Yukawaon model 
in which the phase matrix $P$ appears in the up-quark sector. 
In Sec.3, parameter fitting are discussed.  
Especially,  it is shown 
that $CP$ violating phase 
parameters $\delta_{CP}^q$ and $\delta_{CP}^\ell$ in the standard 
expression of $V_{CKM}$ and $U_{PMNS}$ are predicted as 
$\delta_{CP}^q \simeq 72^\circ$ 
and $\delta_{CP}^\ell \simeq -76^\circ$, respectively, 
i.e. $\delta_{CP}^\ell \sim  - \delta_{CP}^q$.
In Sec.4, we will propose a new relation between $P$ and $m_{ei}$. 
Finally, Sec.5 is devoted to concluding remarks.  
In Appendices A and B, we will demonstrate 
an example of the charged lepton mass relation in the present 
Yukawaon model.

\vspace{5mm}

\noindent{\large\bf 2 \ Model}

\vspace{2mm}

{\bf 2.1 \ VEV relations and superpotentials}

The VEV relations of Yukawaons have already been given in Eqs.(1.3) 
and (1.4) in Sec.1. 
Those VEV relations are derived from a U(3)$\times$U(3)$'$ symmetric 
superpotential. 
For example, for the VEV relation (1.3), we assume \cite{Yukawaon_3}
$$
W_Y = \sum_{f=u,d,\nu,e} \left\{ {\rm Tr}\left[ \left( \mu_f \hat{Y}_f + 
\lambda_f \Phi_f \bar{\Phi}_f \right) \hat{\Theta}_f \right]
+ {\rm Tr}\left[ \mu'_f \hat{Y}_f + \lambda'_f \Phi_f \bar{\Phi}_f \right] 
{\rm Tr}[ \hat{\Theta}_f ] \right\} .
\eqno(2.1)
$$
The supersymmetric vacuum condition $\partial W/ \partial \hat{\Theta}_f =0$
leads to 
$$
\mu_f \hat{Y}_f + \lambda_f \Phi_f \bar{\Phi}_f 
+ {\rm Tr}\left[ \mu'_f \hat{Y}_f + \lambda'_f \Phi_f \bar{\Phi}_f \right] 
{\bf 1} =0 ,
\eqno(2.2)
$$
that is,  to the VEV relation (1.3) with the $\xi_f$ term, 
$$
\xi_f = \frac{ \mu'_f \lambda_f  - \mu_f \lambda'_f}{
(\mu_f + 3 \mu'_f) \mu_f} \, {\rm Tr} [\Phi_f \bar{\Phi}_f ] .
\eqno(2.3)
$$
Here, we assume that flavons $\hat{\Theta}_f$ always take 
$\langle \hat{\Theta}_f \rangle =0$, so that vacuum conditions, which are 
obtained by differentiating the superpotential $W_Y$ with respect to 
other flavons, do not give any additional VEV relations, 
because those relations always include one $\langle \hat{\Theta}_f \rangle$. 

In a similar way, 
we can obtain the VEV relation (1.4) by assuming
$$
W_{\Phi} = \sum_{f=u,d,\nu,e} \left(\lambda_{f1} (\bar{P}_f)^{ik} (\Phi_f)_{kl}  
(\bar{P}_f)^{lj} +\lambda_{f2} ( \bar{\Phi}_0 )^{i\alpha} (S_f)_{\alpha\beta} 
(\bar{\Phi}_0)^{\beta j}  \right) (\Theta_f)_{ij} ,
\eqno(2.4)
$$
and so on. 
Although we assumed a tad pole term 
${\rm Tr}[ \hat{\Theta}_f ]$ in (2.1), we consider that such a term 
is a special case only for $\hat{\Theta}_f$ with $({\bf 8}+{\bf 1})$ 
of U(3).
Therefore, such $\xi_f$ terms do not appear in the relation (1.4). 

In order to give a neutrino mass matrix with seesaw mechanism, 
$M_\nu = k_\nu \langle \hat{Y}_\nu^T \rangle \langle Y_R\rangle^{-1} 
\langle \hat{Y}_\nu \rangle$, correspondingly to the following Majorana 
neutrino mass matrix $M_R$ (i.e. $\langle Y_R \rangle$), we assume 
VEV matrix relation
$$
\langle Y_R \rangle = \langle\hat{Y}_e \rangle \langle\Phi_u \rangle 
+ \langle\Phi_u \rangle \langle\hat{Y}_e^T\rangle , 
\eqno(2.5)
$$
according to the previous Yukawaon model \cite{Yukawaon_3}.


\vspace{2mm}

{\bf 2.2 \ VEV structures of $P_f$}

Prior to discussing VEV forms $\langle {P}_f \rangle$ and 
$\langle \bar{P}_f \rangle$ given in Eq.(1.4),
let us consider the following superpotential
$$
W_P = \frac{\lambda_1}{\Lambda} {\rm Tr}[ P\bar{P} E \bar{E}] +
 \frac{\lambda_2}{\Lambda} {\rm Tr}[ P\bar{P} ] {\rm Tr}[E \bar{E}] ,
 \eqno(2.6)
$$
where, in order to distinguish $P$ from $E$, we assign  $R$ charges of 
$P$ and $E$ as
$$
R(P) =R(\bar{P}) = \frac{1}{2} (1-\Delta), \ \ \ 
R(E) =R(\bar{E}) = \frac{1}{2} (1+\Delta), 
\eqno(2.7)
$$
so that $R(P) +R(\bar{P}) + R(E) +R(\bar{E}) =2$. 
The supersymmetric vacuum conditions lead to 
$$
\langle P \rangle \langle \bar{P} \rangle = {\bf 1} , \ \ \ \ \ 
\langle E \rangle \langle \bar{E} \rangle = {\bf 1} .
\eqno(2.8)
$$
We define specific solutions of (2.8) as 
$$
\langle P\rangle = {\rm diag} (e^{i \phi_1}, e^{i \phi_2}, e^{ i\phi_3} ), \ \ \ 
\langle E \rangle = {\rm diag} (1, 1, 1) .
\eqno(2.9)
$$ 
We consider that VEV of each flavon $\langle P_f \rangle$ 
given in Eq.(1.4) is given by either $\langle P \rangle$ or 
$\langle E\rangle$ in Eq.(2.9) under the $D$-term condition 
as discussed in (2.10) below.  

On the other hand, let us recall that, in general, VEV matrix  
$\langle \bar{A} \rangle$  is related to VEV matrix 
 $\langle {A} \rangle$ under the $D$ term condition as
$$
\langle \bar{A} \rangle = \langle {A} \rangle^* , \ \ \ 
{\rm or} \ \ \ 
\langle \bar{A} \rangle = \langle {A} \rangle.
\eqno(2.10)
$$
Let us back to the relations (1.4).
We take
$$
\langle \Phi_0 \rangle = \langle \bar{\Phi}_0 \rangle = 
{\rm diag}(x_1, x_2. x_3) ,
\eqno(2.11)
$$
$$
\langle S_f \rangle = \langle \bar{S}_f \rangle = 
{\bf 1} + a_f e^{i\alpha_f} X_3 ,
\eqno(2.12)
$$
while
$$
\langle \bar{P} \rangle = \langle P \rangle^* = 
{\rm diag} (e^{-i \phi_1}, e^{- i\phi_2}, e^{-i \phi_3} ) , 
\eqno(2.13)
$$
where parameters $x_i$, $\phi_i$, $a_f$ and $\alpha_f$ are real. 
Then, according as $\langle \bar{\Phi}_f \rangle =\langle {\Phi}_f \rangle^*$
or $\langle \bar{\Phi}_f \rangle =\langle {\Phi}_f \rangle$,
the relations (1.4) require $\langle \bar{P}_f \rangle = \langle {P}_f \rangle^* $
or  $\langle \bar{P}_f \rangle = \langle {P}_f \rangle$. 
For example, when we take the case 
$\langle \bar{\Phi}_f \rangle = \langle {\Phi}_f \rangle^* $,  Eq.(1.4)  becomes
$$
\begin{array}{l}
\langle {P}_f \rangle^*  \langle \Phi_f\rangle \langle {P}_f\rangle^* 
= \langle {\Phi}_0 \rangle  \langle S_f \rangle \langle {\Phi}_0 \rangle , \\
\langle {P}_f\rangle  \langle {\Phi}_f\rangle^* \langle {P}_f\rangle  = 
\langle {\Phi}_0 \rangle \langle {S}_f \rangle \langle {\Phi}_0 \rangle . 
\end{array}
\eqno(2.14)
$$
Two equations in (2.14) cannot simultaneously satisfied without 
$\alpha_f=0$. 
On the other hand, when $\langle \bar{\Phi}_f \rangle =\langle {\Phi}_f \rangle$,
Eq.(1.4) becomes
$$
\begin{array}{l}
\langle {P}_f \rangle^*  \langle \Phi_f\rangle \langle {P}_f\rangle^* 
= \langle {\Phi}_0 \rangle  \langle S_f \rangle \langle {\Phi}_0 \rangle , \\
\langle {P}_f\rangle  \langle {\Phi}_f\rangle \langle {P}_f\rangle  = 
\langle {\Phi}_0 \rangle \langle {S}_f \rangle \langle {\Phi}_0 \rangle .
\end{array}
\eqno(2.15)
$$
Then, for the case of $\alpha_f \neq 0$, the two equations (2.15) are 
satisfied only when $\phi_i=0$ 
[i.e. $\langle P_f \rangle = \langle E \rangle$].
As a result, we consider only two cases: (i) for the case of $\alpha_f \neq 0$,
we regard $\langle P_f \rangle$ as $\langle E \rangle$ given in (2.9), 
and (ii)  only for the case of
$\alpha_f =0$, we regard $\langle P_f \rangle$ as $\langle P \rangle$ 
with $\phi_i \neq 0$. 

The parameters $\alpha_f$ affect not only $CP$ violation, but also mass ratios. 
In the up-quark sector, as we discuss in Sec.3, we can fit up-quark mass 
ratios $m_u/m_c$ and $m_c/m_t$ by taking two parameters $a_u$ 
and $\xi_u$ (keeping $\alpha_u=0$). 
Therefore, we regard the up-quark sector as the case of $\alpha_f=0$,
so that we regard $\langle P_f \rangle$ as 
$\langle P_f \rangle= \langle P \rangle$.  
Also, since we have taken $a_e=0$, we have to regard  $\langle P_e \rangle$ 
as $\langle P_e \rangle = \langle P \rangle$.
Note that $\langle P \rangle$ and $\langle \hat{Y}_e \rangle$ are diagonal, 
so that they are commutable each other. 
Therefore, $\langle P_e  \rangle$ does not play any essential 
physical role in the parameter fitting of the masses and mixing of quarks.  
Hereafter, we denote $\langle P_e  \rangle$ as $\langle E  \rangle$ 
from the practical point of view,
except for a case of counting of $R$ charge. 
On the other hand, in down-quark sector, we cannot fit down-quark 
mass ratios $m_d/m_s$ and $m_s/m_b$ without help of $\alpha_d \neq 0$.
Therefore, we regard down-quark sector as a case of 
$\langle P_f \rangle =\langle E  \rangle$.
Thus, we have the selection rule, $\langle P_f  \rangle =\langle P \rangle$ 
or $\langle P_f  \rangle = \langle E  \rangle$, as a phenomenological one.
For neutrino sector, we have no phenomenological information.
For simplicity, we take a fewer parameter scheme ($\alpha_\nu \neq 0$ 
rather than $\phi_i^\nu \neq 0$).
Hereafter, we will use the notation $P_f$ as 
$$
\begin{array}{ll}
P_f=P \ \ & {\rm for} \ \ f=u, e, \\
P_f=E  \ \ & {\rm for} \ \ f=d, \nu . \\
\end{array}
\eqno(2.16)
$$ 
Sometimes, for convenience, we use notations $P_u$, $P_e$, 
and so on, although we identify $P_u$ and $P_e$ as one flavon $P$, and 
also $P_d$ and $P_\nu$ as one flavon $E$.

The phase matrix $\langle P \rangle$ does not affect mass ratios. 
Since $\langle P_u \rangle =\langle P \rangle$, the phase parameters 
affects $CP$ violation phase
$\delta_{CP}^q$ in the CKM mixing matrix $V_{CKM}$. 
However, note that the phase in $\langle P_u \rangle = \langle P \rangle$ 
can also affect 
$CP$ violation phase $\delta_{CP}^\ell$ in the PMNS mixing matrix
$U_{PMNS}$, because the phase in $P_u$ can affect $Y_R$ 
through $\Phi_u$ as shown in Eq.(2.5). 
This is the most notable point in the present paper.\footnote{
A similar model with $P_u$ has been discussed in 
Ref.\cite{Yukawaon_3}. 
However, we has neglected a possible effect of $\langle P_u$ in the neutrino 
mixing. 
}

The details are discussed in the next section, Sec.3.


\vspace{2mm}

{\bf 2.3 \ $R$ charge assignments}

In this model, the number of flavons is larger than 
that of VEV relations.
Therefore, in general, we cannot uniquely determine $R$ charges of flavons.
Since we  demand to assign $R$ charges as simple as possible, we put 
the following rules for simplicity: 

\noindent (i) We assign the same $R$ charge to flavons $A$ and $\bar{A}$: 
$$
R(A) = R(\bar{A}) ,
\eqno(2.17)
$$
independently 
whether $\langle \bar{A} \rangle = \langle {A} \rangle^*$ or 
$\langle \bar{A} \rangle = \langle {A} \rangle$.  
Then, we obtain $R$ charge relations
$$
R(\hat{Y}_f) = 2 R(\Phi_f) \equiv 2 r_f  \ \ \  (f=u, d, \nu, e) , 
\eqno(2.18)
$$
and
$$
R(\Phi_f) = R(\bar{\Phi}_f) = R(S_f) +2 R(\Phi_0) -2 R(P_f) 
\ \ \ \ (f=u, d, \nu, e),
\eqno(2.19)
$$
from Eqs.(1.3) and (1.4), and 
$$
\begin{array}{l}
R(P_u) =R(P_e)= R(P) \equiv \frac{1}{2}(1+\Delta) , \\
R(P_d) =R(P_\nu)= R(E) \equiv \frac{1}{2}(1-\Delta) , 
\end{array}
\eqno(2.20)
$$
from Eqs.(2.16) and (2.7).  
Therefore, from Eq.(2.19), we obtain the following relations:
$$
\begin{array}{l}
R(\Phi_e) -R(S_e) = R(\Phi_u) -R(S_u) = 2 R(\Phi_0) -(1+\Delta) , \\
R(\Phi_\nu) -R(S_\nu) = R(\Phi_d) -R(S_d) = 2 R(\Phi_0) -(1-\Delta) . \\
\end{array}
\eqno(2.21)
$$

\noindent (ii) We can regard that $R$ charges of $\hat{Y}_f$ are determined 
only by those of the SU(2)$_L$ singlet fermions $f^c$.
Therefore, we simply assign 
$$
R(\ell H_u) = R(\ell H_d) = R(q H_u) = R(q H_d) \equiv r_H +2.
\eqno(2.22)
$$
Since those have different quantum number of U(1)$_Y$, 
we can distinguish those from each other in spite of the
relation (2.22). 
Then, we obtain a simple $R$ charge relation
$$
R(\hat{Y}_f) + R(f^c) = -r_H .
\eqno(2.23)
$$

For $Y_R$, we obtain
$$
R(Y_R) = 2-2R(\nu^c) =  2r_H +2 +2 R(\hat{Y}_\nu) ,
\eqno(2.24)
$$
from Eqs.(2.1) and (2.23). 
On the other hand, from Eq.(2.5), $R(Y_R)$ must be satisfied a relation 
$$
R(Y_R) = R(\Phi_u) + 2 R(\Phi_e) .
\eqno(2.25)
$$
From Eqs.(2.24) and (2.25), we have the following constraint
$$
2 R(\Phi_e) - 4 R(\Phi_\nu) + R(\Phi_u) = 2 r_H +2 .
\eqno(2.26)
$$


Even under the these constraints, we cannot still completely 
fix the $R$ charges of whole flavons.
In the present model, $R$ charge assignments are not  so essential, 
so that it is enough to assign R charges to distinguish flavons with the same U(3) 
from each other.  
That is, we are satisfied with  any $R$-charge numbers which satisfy 
the relations (2.18) - (2.26). 
Nevertheless, it is desirable to have explicit $R$-charge assignments as simple as 
possible.
Therefore, let us go on our search for explicit $R$-charge assignments.  

First, for simplicity, we put
$$
R(\Phi_0) = \frac{1}{2} .
\eqno(2.27)
$$
Then. Eq.(2.21) becomes to be simpler relations
$$
\begin{array}{l}
R(S_f) = R(\Phi_f) + \Delta \ \ \ (f=e, u) , \\
R(S_f) = R(\Phi_f) - \Delta \ \ \ (f=\nu, d) . 
\end{array}
\eqno(2.28)
$$

Now, let us discuss possible $R$-charge assignments for Yukawaons $\hat{Y}_f$ under 
the conditions discussed above.
If we have $R(\hat{Y}_f)=0$, then we can attach the field $\hat{Y}_f$ on 
any term in superpotential. 
Therefore, we require $R(\hat{Y}_f) \neq 0$ for any $f=e, \nu, d, u$.
Also, we have to require  $R(\hat{Y}_f \hat{Y}_{f'}) \neq 0$ for any combination
of $f$ and $f'$. 
As a result, we have to consider that whole $R$ values of $\hat{Y}_f$  
are positive. 
Furthermore, we speculate that the values of $R$ will be describe by simple integers.
Of course, the $R$ charges have to satisfy the relation (2.26). 
Therefore, we assign simpler $R$ charges to the Yukawaons $\hat{Y}_f$ on trial as follows:
$$
\left( R(\hat{Y}_e), R(\hat{Y}_u), R(\hat{Y}_\nu),  R(\hat{Y}_d) \right) 
= (1, 2, 3, 4), 
\eqno(2.29)
$$
that is, 
$$
\left( R(\Phi_e), R(\Phi_u), R(\Phi_\nu),  R(\Phi_d) \right) 
= \left( \frac{1}{2},1,\frac{3}{2}, 2 \right). 
\eqno(2.30)
$$
This assignment satisfies the condition (2.25) for $R(Y_R)$ with $r_H=-3$.

In Table~1 , as a summary of Sec.2, we present the assignments of 
SU(2)$_L \times$SU(3)$_c \times$U(3)$\times$U(3)$'$ and 
the $R$ charges of the fields in the present model.


\begin{table}
\caption{
Assignments of SU(2)$_L \times$SU(3)$_c \times$U(3)$\times$U(3)$'$. 
For $R$ charges, see subsection 2.3. 
We assign the same $R$ charges   
for flavons $A$ and $\bar{A}$, e.g. $R(A)=R(\bar{A})$. 
For a special choice, $r_e$, $r_\nu$, 
$r_u$ and $r_d$ are taken as $r_e=1/2$, $r_\nu=3/2$, 
$r_u=2/2$ and  $r_d=4/2$. 
}

\begin{center}
\begin{tabular}{|c|cc|cc|cc|} \hline
& $\ell = (\nu, e)$ & $f^c=\nu^c, e^c$  & 
$q=(u, d)$ & $f^c=u^c, d^c$ & 
$H_u$ & $H_d$ 
\\ \hline
SU(2)$_L$ & ${\bf 2}$ & ${\bf 1}$ & ${\bf 2}$ & ${\bf 1}$  & 
${\bf 2}$ & ${\bf 2}$ 
\\ 
SU(3)$_c$ & ${\bf 1}$ & ${\bf 1}$ & ${\bf 3}$ & ${\bf 3}^*$ &
${\bf 1}$ & ${\bf 1}$  \\ \hline
U(3) & ${\bf 3}$ & ${\bf 3}^*$ & ${\bf 3}$ & ${\bf 3}^*$ & 
 ${\bf 1}$ & ${\bf 1}$  \\
U(3)$'$ & ${\bf 1}$ & ${\bf 1}$ & ${\bf 1}$ & ${\bf 1}$ & 
 ${\bf 1}$ & ${\bf 1}$  
\\  \hline
$R$     & 2 & $-(2 r_{f}+r_H)$  & 2 & $-(2 r_{f}+r_H)$ & 
$r_H$  & $r_H$ 
\\  \hline 
\end{tabular}

\vspace{2mm}

\begin{tabular}{|cc|cc|cc|cc|cc|} \hline
$\hat{Y}_f$ & ${Y}_R$ & 
$\bar{\Phi}_f$ & $\Phi_f$ & $\bar{\Phi}_0$ & $\bar{\Phi}_0$ &
$S_{e,u}$ & $\bar{S}_{e,u}$ & $S_{\nu, d}$ & $\bar{S}_{\nu, d}$ 
\\ \hline
 ${\bf 1}$ & ${\bf 1}$ & ${\bf 1}$ & ${\bf 1}$ &  ${\bf 1}$ & 
${\bf 1}$ &  ${\bf 1}$ &  ${\bf 1}$ & ${\bf 1}$ &  ${\bf 1}$
\\ 
 ${\bf 1}$ & ${\bf 1}$ & ${\bf 1}$ & ${\bf 1}$ &  ${\bf 1}$ & ${\bf 1}$ &
 ${\bf 1}$ &  ${\bf 1}$ &  ${\bf 1}$ &  ${\bf 1}$
\\ \hline
 ${\bf 8}+{\bf 1}$ & ${\bf 6}$ &  ${\bf 6}$ & ${\bf 6}^*$ &  
 ${\bf 3}$ & ${\bf 3}^*$ &  ${\bf 1}$ &  ${\bf 1}$ &  ${\bf 1}$ &  ${\bf 1}$
\\ 
 ${\bf 1}$ & ${\bf 1}$ &  ${\bf 1}$ & ${\bf 1}$ & 
${\bf 3}$ & ${\bf 3}^*$ &  ${\bf 6}$ &  ${\bf 6}^*$ & ${\bf 6}$ &  ${\bf 6}^*$
\\ \hline
$2r_f$ & $r_R$ & 
\multicolumn{2}{|c}{$r_f$} & \multicolumn{2}{|c|}{$1/2$}  &
\multicolumn{2}{|c|}{$ r_{e,u} -\Delta$} & 
\multicolumn{2}{|c|}{$ r_{\nu, d}  + \Delta $} 
\\ \hline
\end{tabular}

\vspace{2mm}

\begin{tabular}{|cccc|c|c|cc|} \hline
${P}$ & $\bar{P}$ & ${E}$ & $\bar{E}$ &
$\hat{\Theta}_f$ & $\bar{\Theta}_R$ & 
$\Theta_{\Phi f}$ & $\bar{\Theta}_{\Phi f}$  
\\ \hline
  ${\bf 1}$ &  ${\bf 1}$ &  ${\bf 1}$ &  ${\bf 1}$ & 
${\bf 1}$ &  ${\bf 1}$ & ${\bf 1}$ &  ${\bf 1}$
\\ 
  ${\bf 1}$ &  ${\bf 1}$ & ${\bf 1}$ & ${\bf 1}$ &  
 ${\bf 1}$ & ${\bf 1}$ & ${\bf 1}$ &  ${\bf 1}$
\\ \hline
 ${\bf 6}$ & ${\bf 6}^*$ & ${\bf 6}$ & ${\bf 6}^*$ & 
${\bf 8}+{\bf 1}$ &  ${\bf 6}^*$ &  ${\bf 6}$ & ${\bf 6}^*$ 
\\
${\bf 1}$ &  ${\bf 1}$ & ${\bf 1}$ &  ${\bf 1}$ & 
${\bf 1}$ & ${\bf 1}$ & ${\bf 1}$ & ${\bf 1}$
\\ \hline
\multicolumn{2}{|c}{$\frac{1}{2}(1+\Delta)$}  & 
\multicolumn{2}{c|}{$\frac{1}{2}(1-\Delta)$}  & 
 $2-2r_f$   & $2- r_R$  & \multicolumn{2}{|c|}{$1-R(S_f)$}      
\\ \hline
\end{tabular}

\end{center}

\end{table}

\vspace{5mm}

\noindent{\large\bf 3 \ Parameter fitting}

\vspace{2mm}

\noindent{\bf 3.1 \ How many parameters?}

Our mass matrices $Y_f $  for $f=e, \nu, d, u$  with the VEV relations discussed in Sec.1 and Sec.2 are summarized as follows:
$$
\begin{array}{ll}
Y_e  = \Phi_e \Phi_e^* , & \\ 
 & \Phi_e = P^* \Phi_0 \Phi_0 P^*, \\
 & \Phi_e^* = P \Phi_0 \Phi_0 P, \\ 
 & \Phi_0 = {\rm diag}(x_1, x_2, x_3), \\
 \end{array}
\eqno(3.1)
$$
$$
\begin{array}{ll}
Y_\nu = \Phi_\nu \Phi_\nu + \xi_\nu {\bf 1} , &  \\ 
 &  \Phi_\nu = E\, \Phi_0 ({\bf 1}+ a_\nu e^{i\alpha_\nu} X_3) \Phi_0 E , \\
 \end{array}
 \eqno(3.2)
$$
$$
\begin{array}{ll}
Y_u = \Phi_u \Phi_u^* + \xi_u {\bf 1} , \\
 &  \Phi_u = P^* \Phi_0 ({\bf 1} +a_u X_3) \Phi_0 P^* , \\
 &  \Phi_u^* = P \Phi_0 ({\bf 1} +a_u X_3) \Phi_0 P  , \\ 
 \end{array}
 \eqno(3.3)
$$
$$
\begin{array}{ll}
Y_d = \Phi_d \Phi_d + \xi_d {\bf 1} , &  \\  
 &  \Phi_d = E\, \Phi_0 ({\bf 1}+ a_d e^{i\alpha_d} X_3) \Phi_0 E , \\
 \end{array}
 \eqno(3.4)
$$
Neutrino mass matrix with seesaw mechanism is given by
$$
\begin{array}{ll}
M_\nu = Y_\nu Y_R^{-1} Y_\nu , & \\ 
  & Y_R = Y_e \Phi_u  + \Phi_u Y_e , \\
\end{array}
\eqno(3.5)
$$
Note that $\Phi_e^*$ in Eq.(3.1) and $\Phi_u^*$ in Eq.(3.3) are 
not $\Phi_e$ and $\Phi_u$, respectively. 
Here, for convenience, we have dropped 
the notations ``$\langle$" and ``$\rangle$". 
We also make no distinction of  property under U(3)$\times$U(3)$'$, 
i.e. we denote $\hat{A}$ and also $\bar{A}$ as $A$ simply.    
Since we are interested only in the mass ratios and mixings, 
we use dimensionless expressions
$\Phi_0 = {\rm diag}(x_1, x_2, x_3)$ (with $x_1^2+x_2^2+x_3^2=1$),  
$P= {\rm diag} (e^{i\phi_1}, e^{i\phi_2},1)$, 
and $E={\bf 1}={\rm diag}(1,1,1)$. 
Therefore, the parameters $a_e$, $a_\nu$, $a_u$, $a_d$, $\xi_\nu$, $\xi_u$, 
and $\xi_d$ are re-defined by Eqs.(3.1)-(3.4).

In the phase matrix $P$ defined by Eq.(1.10), 
physical values are only differences among $(\phi_1, \phi_2, \phi_3)$, 
so that we can take one of $\phi_i$ ($i=1,2,3$) as zero in the 
parameter fitting for $V_{CKM}$. 
In this paper, we put $\phi_3=0$, so that free parameters are
$(\phi_1, \phi_2)$. 
Note that, as we stated in Sec.2.2, $P$ and $P^*$ in Eq.(3.1) do not 
affect $Y_e$ practically,
because $\Phi_0$ and $Y_e$ are diagonal, so that $P$ and $P^*$ are 
commutable with $\Phi_0$ and $Y_e$.

Therefore, in the present model shown in Eqs.(3.1)  - (3.5), except for the parameters 
$(x_1, x_2, x_3)$, we have 10 adjustable parameters,  
$(a_\nu, \alpha_\nu, \xi_\nu)$, $(a_u, \xi_u)$, $(a_d, \alpha_d,  \xi_d)$,
 and $(\phi_1, \phi_2)$   
for the 16 observable quantities (6 mass ratios in the
up-quark, down-quark, and neutrino sectors, four CKM 
mixing parameters, and 4+2 PMNS mixing parameters). 
Especially, quark mass matrices $M_u=Y_u$ and $M_d=Y_d$ are fixed
by two parameters $(a_u, \xi_u)$ and $(a_d, \alpha_d, \xi_d)$, 
respectively.
Note that those parameters are family number-independent 
parameters.
Therefore, in order to fix those parameters, we use two inputs
values, up-quark mass ratios $(m_u/m_c, m_c/m_t)$ and 
down-quark mass ratios $(m_d/m_s, m_s/m_b)$, respectively,
as we discuss in the next subsection 3.2.
After the parameters $(a_u, \xi_u)$ and $(a_d, \alpha_d, \xi_d)$
have been fixed by the observed quark mass rations, we have
five parameters $(a_\nu, \alpha_\nu, \xi_\nu)$ and $(\phi_1, \phi_2)$
as remaining free parameters. Processes for fitting those five parameters 
are listed in Table~2. 
In subsection 3.3, we discuss the fitting of four CKM mixing parameters,  
$|V_{us}|$, $|V_{cb}|$, $|V_{ub}|$ and $|V_{td}|$, by adjusting 
two parameters $(\phi_1, \phi_2)$. 
Also, in subsection 3.4, we do the fitting of PMNS mixing ($\sin^2 2 \theta_{12}$, 
$\sin^2 2 \theta_{23}$, and $\sin^2 2 \theta_{13}$) 
and neutrino mass ratio 
($R_\nu \equiv {\Delta m_{21}^2}/{\Delta m_{32}^2}$) 
 by adjusting three parameters $(a_\nu, \alpha_\nu, \xi_\nu)$.

\vspace{2mm}

\begin{table}
\caption{Process for fitting parameters. 
 $N_{parameter}$ and $N_{input}$ denote a number of free 
parameters in the model and a number of observed values which
are used as inputs in order to fix these free parameters, 
respectively. $\sum N_{\dots}$ means $\sum N_{parameter}$ or  $\sum N_{input}$
}
\vspace{2mm}
\begin{center}
\begin{tabular}{|c|cc|cc|c|} \hline
Step & Inputs & $N_{input}$ &  Parameters & $N_{parameter}$ &
 Predictions  \\ \hline
 1st   & $m_u/m_c$, $m_c/m_t$ & 2 & $a_u$, $\xi_u$ & 2 & --- \\
     &  $m_d/m_s$, $m_s/m_b$ & 2 & $a_d$, $\alpha_d$, $\xi_d$ & 3 & --- \\
2nd  & $|V_{cb}|$, $|V_{ub}|$  & 2 &   $(\phi_1, \phi_2)$ & 2 & 
$|V_{us}|$,  $|V_{td}|$, $\delta_{CP}^q$  \\
3rd  & $\sin^2 2\theta_{12}$, $\sin^2 2\theta_{23}$, $R_\nu$  & 3 
&  $a_\nu$, $\alpha_\nu$, $\xi_\nu$  & 3 & 
$\sin^2 2\theta_{13}$, $\delta_{CP}^\ell$  \\
    &    &   &   &  & 2 Majorana phases, $\frac{m_{\nu 1}}{m_{\nu 2}}$,
 $\frac{m_{\nu 2}}{m_{\nu 3}}$   \\ 

option &  $\Delta m^2_{32}$ &   & $m_{\nu 3}$ &  & 
$(m_{\nu 1}, m_{\nu 2},  m_{\nu 3})$, $\langle m \rangle$  \\
\hline
$\sum N_{\dots}$ & & 9 &  & 10 &   \\ \hline 
\end{tabular}
\end{center}
\end{table}

\vspace{2mm}

\noindent{\bf 3.2 \ Quark mass ratios}

First let us fix values of $(a_u, \xi_u)$ from the up-quark mass ratios. 
The observed values of the up-quark masses at $\mu=m_Z$ \cite{q-mass} are 
$$
r^u_{12} \equiv \sqrt{\frac{m_u}{m_c}} 
= 0.045^{+0.013}_{-0.010} , \ \ \ \ 
r^u_{23} \equiv \sqrt{\frac{m_c}{m_t}}
=0.060 \pm 0.005 .
\eqno(3.6)
$$ 
We obtain four solutions of $(a_u, \xi_u)$ which can give
the values (3.6). 
Among them only one solution
$$
(a_u, \xi_u) = (- 1.4715, -0.001521) ,
\eqno(3.7)
$$
can give a reasonable prediction of the PMNS mixing
as we discuss later.  

Secondly, let us fix values of $(a_d, \alpha_d, \xi_d)$ from the down-quark mass ratios. 
From the observed down-quark mass ratios \cite{q-mass}
$$
r^d_{12} \equiv \frac{m_d}{m_s} = 0.053^{+0.005}_{-0.003} , \ \ \ 
r^d_{23} \equiv \frac{m_s}{m_b} = 0.019 \pm 0.006 , 
\eqno(3.8)
$$
or \cite{q-mass2}
$$
r^d_{12} \equiv \frac{m_d}{m_s} = 0.050^{+0.002}_{-0.001} , \ \ \ 
r^d_{23} \equiv \frac{m_s}{m_b} = 0.031 \pm 0.005 , 
\eqno(3.9)
$$
we determine the parameters $(a_d, \xi_d, \alpha_d)$ as follows: 
$$
(a_d, \alpha_d, \xi_d) = (- 1.4733,15.694^\circ, +0.004015),
\eqno(3.10)
$$
which leads to the numerical results as follows: 
$r^d_{12}=0.0612$, $r^d_{23}=0.0312$. 
These values are inconsistent with the observed values (3.9), 
but, roughly speaking, those are consistent with (3.10).
We think that the light quark mass values are still controversial. 

Here, we have tried to fix the parameters $(a_d, \alpha_d, \xi_d)$ in 
the down-quark sector by using input parameters \cite{q-mass} for 
$r^d_{12} $ and $r^d_{23}$.
However, since we have three parameters for two input values
$m_d/m_s$ and $m_s/m_b$, we cannot fix our three parameters.
It is more embarrassing that there is no solution of 
$m_s/m_b \sim 0.019$ in the $(a_d, \alpha_d, \xi_d)$ parameter region. 
Nevertheless, we found that the minimal value of $m_s/m_b$ is $m_s/m_b\sim 0.03$ at 
 $(a_d, \alpha_d, \xi_d)\sim (-1.47, 16^\circ, 0.004)$ which
can give a reasonable value of $m_d/m_s$ at the same time too.
Therefore, we take the values in Eq.~(3.10), which leads to 
$r^d_{12}=0.0612$ and $r^d_{23}=0.0312$.
Note that the value $r_{23}^d = 0.0312$ is considerably
large compared with $r_{23}^d \simeq 0.019$ by Xing {\it et al.}
\cite{q-mass}, 
while the value is consistent with  $r_{23}^d \simeq 0.031$ by 
Fusaoka and Koide \cite{q-mass2}.
The values $m_d(\mu)$ and $m_s(\mu)$ are estimated at a lower energy scale, 
$\mu \sim 1$ GeV, so that we consider that the ratio $r_{12}^d$ 
at $\mu = M_Z$ is reliable. 
On the other hand, the value $m_b(\mu)$ is extracted at a different 
energy scale $\mu \sim 4$ GeV from $\mu \sim 1$ GeV, so that 
the value $m_b(M_Z)$ is affected by the prescription of threshold effects 
at $\mu=m_t$, while the value $m_s(M_Z)$ affected by those at $\mu=m_c$,  
$\mu=m_b$ and $\mu = m_t$.
We consider that as for the ratio $r_{23}^d$ at $\mu = M_Z$ 
the value is still controversial.
Anyhow, we have fixed three parameters $(a_d, \alpha_d, \xi_d)$ only
from two values $m_d/m_s$ and $m_s/m_b$.

\vspace{2mm}

\noindent{\bf 3.3 \ CKM mixing}  

Next, we discuss CKM quark mixing. 
Since the parameters $(a_u, \xi_u)$ and $(a_d, \alpha_d, \xi_d)$
have been fixed by the observed quark mass rations, 
the CKM mixing matrix elements  $|V_{us}|$, 
$|V_{cb}|$, $|V_{ub}|$, and  $|V_{td}|$ are functions of 
the remaining two parameters 
$\phi_1$ and $\phi_2$.
In Fig.~2, with taking  $\xi_u=-0.001521$, $a_u=-1.4715$, $a_d=-1.47312$, 
$\alpha_d=15.7^\circ$, and $\xi_d=0.004091$, we draw allowed regions 
in the ($\phi_1$, $\phi_2$) 
parameter plane which are obtained from the observed values  \cite{PDG14} of 
the CKM mixing matrix elements  and  the observed value \cite{UTfit}  of the $CP$ violating phase 
parameter $\delta_{CP}^q$ in the standard 
expression of $V_{CKM}$  given by, 
$$
\begin{array}{l}
|V_{us}|=0.22536 \pm 0.00061, \\
|V_{cb}|=0.0414 \pm 0.0012,  \\
|V_{ub}|=0.00355 \pm 0.00015,  \\
|V_{td}|=0.00886^{+0.00033}_{-0.00032} ,\\
\delta_{CP}^q=69.4^\circ \pm 3.4^\circ .\\
\end{array}
\eqno(3.11)
$$
Here, in order to fix  the values of  ($\phi_1$, $\phi_2$) we use 
only two values of the CKM matrix elements as input values in the present 
analysis, so that the remaining tree are our predictions as references.\\

\begin{figure}[ht]
\begin{picture}(200,200)(0,0)

  \includegraphics[height=.33\textheight]{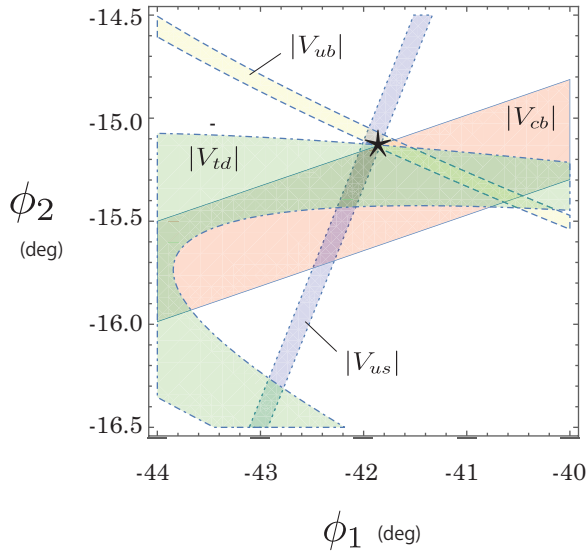}

\end{picture}  
  \caption{Allowed region in the ($\phi_1$, $\phi_2$) parameter plane obtained 
by the observed values of the CKM mixing matrix elements  $|V_{us}|$, 
$|V_{cb}|$, $|V_{ub}|$, and  $|V_{td}|$. 
We draw allowed regions obtained from the observed constraints of the CKM mixing 
matrix elements shown in Eq.~(3.11), 
with taking  $(a_u, \xi_u) = (- 1.4715, -0.001521) $ and 
$(a_d, \alpha_d, \xi_d) = (- 1.4733,15.694^\circ, +0.004015)$ .
We find that the parameter set around 
($\phi_1$, $\phi_2$) $=(-41.815^\circ, -15.128^\circ )$ 
indicated by a star ($\star$) is 
consistent with all the observed values.} 
 \label{fig1}
\end{figure}

As shown in Fig.~1, all the experimental constraints on 
CKM parameters are satisfied by 
fine tuning the parameters $\phi_1$ and $\phi_2$ as
$$
(\phi_1, \phi_2)=(-41.815^\circ, -15.128^\circ ),
\eqno(3.12)
$$
which leads to the predicted values for the CKM mixing matrix elements  and  the  $CP$ violating phase 
parameter $\delta_{CP}^q$ as follows: 
$$
\begin{array}{l}
|V_{us}|=0.2261, \\
|V_{cb}|=0.0426,  \\
|V_{ub}|=0.00360,  \\
|V_{td}|=0.00920 ,\\
\delta_{CP}^q=72.4^\circ .\\
\end{array}
\eqno(3.13)
$$

In spite of our aim described in the Sec.~1, we are forced to 
introduce family number-dependent parameters $(\phi_1, \phi_2)$ 
in the present model, too,  
as the same as in the previous model \cite{Yukawaon_3}. 
However, our aim was to describe all the masses and mixing of quarks and leptons
in terms of family number-independent parameters expect for the charged 
lepton masses. Therefore, the introduction of 
the phase parameters $\phi_i$ were  against our aim and unwelcome as it is.
In Sec.4, we will try to denote these phase parameters $\phi_i$  in terms of  the observed charged lepton masses $m_{ei}$.

\vspace{2mm}

\noindent{\bf 3.4 \ PMNS mixing} 

Now let us discuss the PMNS lepton mixing. 
We have already fixed the four parameters  $a_u$, $\xi_u$,  $\phi_1$ and 
$\phi_2$ as Eqs.~(3.7) and (3.12).
The remaining free parameters in the neutrino sector are only $(a_\nu, \alpha_\nu, \xi_\nu)$.
We determine the parameter values of $(a_\nu, \alpha_\nu, \xi_\nu)$ 
as follows:
$$
(a_\nu, \alpha_\nu, \xi_\nu) = ( -2.59, -27.3^\circ, -0.0115),
\eqno(3.14)
$$
which are obtained 
so as to reproduce the observed values \cite{PDG14}  of the following 
PMNS mixing angles and $R_{\nu}$,
$$
\sin^2 2\theta_{12} = 0.846 \pm 0.021, \ \ \ 
\sin^2 2\theta_{23} >0.981, \ \ \ 
\sin^2 2\theta_{13} = 0.093 \pm 0.008, \\
\eqno(3.15)
$$
$$
R_{\nu} \equiv \frac{\Delta m_{21}^2}{\Delta m_{32}^2}
=\frac{m_{\nu2}^2 -m_{\nu1}^2}{m_{\nu3}^2 -m_{\nu2}^2}
=\frac{(7.53\pm 0.18) \times 10^{-5}\ {\rm eV}^2}{
(2.44 \pm 0.06) \times 10^{-3}\ {\rm eV}^2} = 
(3.09 \pm 0.15) \times 10^{-2} .
\eqno(3.16)
$$

\begin{figure}[ht]
\begin{picture}(200,200)(0,0)
\includegraphics[height=.3\textheight]{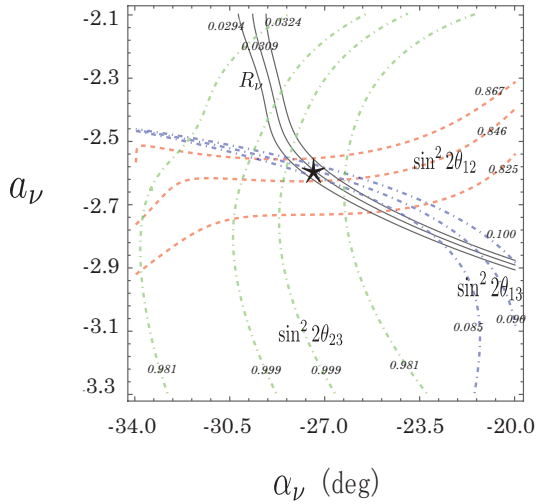}
\end{picture}  
 \caption{ Contour curves of the center, upper, and lower values  
of the observed PMNS mixing parameters 
$\sin^2 2\theta_{12}$, $\sin^2 2\theta_{23}$, $\sin^2 2\theta_{13}$, 
and $R_{\nu}$  in the ($a_\nu$, $\alpha_\nu$) parameter space.
We draw the curves for the case of $\xi_\nu=-0.0115$ and 
$(\phi_1, \phi_2)=(-41.815^\circ, -15.128^\circ )$ 
with taking $(a_u, \xi_u) = (- 1.4715, -0.001521) $. 
We find that the parameter set around $(a_\nu, \alpha_\nu) = ( -2.59, -27.3^\circ)$ 
indicated by a star ($\star$) is 
consistent with all the observed values.
}
\label{fig2}
\end{figure}

In Fig.~2, we show the contour plots of the observed PMNS mixing parameters 
$\sin^2 2\theta_{12}$, $\sin^2 2\theta_{23}$, $\sin^2 2\theta_{13}$, 
and $R_{\nu}$  in the ($a_\nu$, $\alpha_\nu$) parameter space 
for the case of $\xi_\nu=-0.0115$ with taking  
$(\phi_1, \phi_2)=(-41.815^\circ, -15.128^\circ )$ and 
$(a_u, \xi_u) = (- 1.4715, -0.001521)$. 
It is found that all the PMNS mixing parameters are well consistent 
with the observe values in Eqs.~(3.15) and (3.16).
As shown in Fig.~2, all the experimental constraints on 
the PMNS mixing parameters are satisfied by 
fine tuning the parameters $a_\nu$,  $\alpha_\nu$, and $\xi_\nu$ as
$$
(a_\nu, \alpha_\nu, \xi_\nu) = ( -2.59, -27.3^\circ, -0.0115),
\eqno(3.17)
$$
which leads to the predicted values for the PMNS mixing angles,  $R_{\nu}$,  
and  the  Dirac $CP$ violating phase 
parameter $\delta_{CP}^\ell$  in the standard 
expression of $U_{PMNS}$ as follows: 
$$
\begin{array}{l}
\sin^2 2\theta_{12}=0.857, \\
\sin^2 2\theta_{23}=0.993,  \\
\sin^2 2\theta_{13}=0.0964,  \\
R_{\nu}=0.0316 ,\\
\delta_{CP}^\ell=-76.0^\circ .\\
\end{array}
\eqno(3.18)
$$
Our model predicts $\delta_{CP}^{\ell}= -76.0^\circ$ for the Dirac $CP$ 
violating phase in the lepton sector. 
This is very interesting because the value shows a size similar to  
$\delta^q_{CP}= + 72.4^\circ$ in the CKM mixing matrix.
\vspace{2mm}

\noindent{\bf 3.5 \ Neutrino masses}  

We can predict neutrino masses, for the parameters given by (3.7), (3.12), 
and (3.14), as follows
$$
m_{\nu 1} \simeq 0.00046\ {\rm eV}, \ \ m_{\nu 2} \simeq 0.00879 \ {\rm eV}, 
\ \ m_{\nu 3} \simeq 0.0502 \ {\rm eV}  ,
\eqno(3.19)
$$
by using the input value \cite{PDG14}
$\Delta m^2_{32}\simeq 0.00244$ eV$^2$.

We also predict the effective Majorana neutrino mass \cite{Doi1981} 
$\langle m \rangle$ 
in the neutrinoless double beta decay as
$$
\langle m \rangle =\left|m_{\nu 1} (U_{e1})^2 +m_{\nu 2} 
(U_{e2})^2 +m_{\nu 3} (U_{e3})^2\right| 
\simeq 3.8 \times 10^{-3}\ {\rm eV}.
\eqno(3.20)
$$

In Table~3, we list our predictions of the CKM and the PMNS mixing parameters and quark mass ratios  and neutrino masses  together with the observed values .

\begin{table}
\caption{Predicted values vs. observed values. 
} 

\vspace*{2mm}
\hspace*{-6mm}
\begin{tabular}{|c|ccccccccc|} \hline
  & $|V_{us}|$ & $|V_{cb}|$ & $|V_{ub}|$ & $|V_{td}|$ & 
$\delta^q_{CP}$ &  $r^u_{12}$ & $r^u_{23}$ & $r^d_{12}$ & $r^d_{23}$ 
 \\ \hline 
Pred &$0.2261$ & $0.0426$ & $0.00360$ & $0.00920$ & $72.4^\circ$ & 
$0.0458$ & $0.0600$ & $0.0611$ & $0.0312$ 
 \\
Obs & $0.22536$ & $0.0414$ &  $0.00355$  & $0.00886$  & $69.4^\circ$ &
$0.045$ & $0.060$ & $0.053$  & $0.019$  
  \\ 
    &  $ \pm 0.00061$ &  $ \pm 0.0012$& $ \pm 0.00015$ & 
 ${}^{+0.00033}_{-0.00032}$ & $\pm 3.4^\circ$ &
${}^{+0.013}_{-0.010}$ & $ \pm 0.005$ & $^{+0.005}_{-0.003}$ &
${}^{+0.006}_{-0.006}$ 
 \\ \hline
   & $\sin^2 2\theta_{12}$ & $\sin^2 2\theta_{23}$ & $\sin^2 2\theta_{13}$ & 
 $R_{\nu}\ [10^{-2}]$ &  
$\delta^\ell_{CP}$ & $m_{\nu 1}\ [{\rm eV}]$ & $m_{\nu 2}\ [{\rm eV}]$ & 
$m_{\nu 3}\ [{\rm eV}]$ & $\langle m \rangle \ [{\rm eV}]$ \\ \hline
 Pred & $0.857$ & $0.993$ & $0.0964$ &  $3.16$ & $-76.0^\circ$ &
 $0.00046$ & $0.00879$ & $0.0502$ & $0.00377$ \\
Obs & $0.846$   & $ 0.999$ & $0.093$ &   $3.09 $    & -
  &  -  &  -  &  -  &  $<\mathrm{O}(10^{-1})$   \\ 
    & $ \pm 0.021$ & $^{+0.001}_{-0.018}$   & $\pm0.008$ & $ \pm 0.15 $  &  &
   &    &    &    \\ \hline 
\end{tabular}
\end{table}

\vspace{5mm}

\noindent{\large\bf 4 \ VEV relation between $P$ and $\Phi_0$}

So far, we have tried to described all Yukawaon VEV matrices 
$\langle \hat{Y}_f \rangle$ by using only the observed charged lepton 
masses $m_{ei}$ as input values. 
We have also tried to understand $CP$ violating phase only by using phase  
parameters $\alpha_f$ which are phases of family number-independent 
parameters $a_f$.
Nevertheless, all such attempts have failed because 
we always needed a phase matrix $P$ in order to fit reasonable CKM 
mixing and quark mass ratios. 
In this paper, we accept the existence of $P$, and we try to understand 
the values of the phase parameters $\phi_i$  in $P$ 
from the charged lepton mass values $m_{ei}$. 

In the present model, we have flavon VEVs with diagonal form, $P$, $\bar{P}$, 
$E$, $\bar{E}$, $\Phi_0$, $\bar{\Phi}_0$,   $\Phi_e$, $\bar{\Phi}_e$, 
and $\hat{Y}_e$. 
(Here, we omit ``$\langle$" and ``$\rangle$" .)
In considering combinations of U(3) ${\bf 8}+{\bf 1}$ scalars out of those flavons,
we have to consider a combination without 
the parameter $\Delta$ for $E$ and $P$ because the $R$ charges of $\Phi_0$ 
and  $\Phi_e$ do not contain the 
parameter $\Delta$. 
Only a combination with $P$ whose $R$ charge does not include the parameter
$\Delta$ is 
$$
(P \bar{E} +E \bar{P})_i^j = \delta_i^j
(e^{i\phi_i} + e^{-i\phi_i}) =\delta_i^j 2 \cos \phi_i ,
\eqno(4.1)
$$
with $R$ charge of $R=\frac{1}{2}(1+\Delta)  +\frac{1}{2}(1-\Delta) =+1$.
On the other hand, since we have $R$ charges
$$
R(\Phi_e) = \frac{1}{2} , \ \ \ \  R(\Phi_0) = \frac{1}{2} ,
\eqno(4.2)
$$
for $\Phi_e$ and $\Phi_0$ as discussed in Sec.2.3, 
we have only two combinations which have $R$ charge of $R=+1$,   
$(\Phi_e)_{ik} (\bar{\Phi}_e)^{kj}$ and 
$(\Phi_0)_{i\alpha} (\bar{\Phi}_0)^{\alpha j}$.
[Note that $(\Phi_0 \bar{\Phi}_e +\Phi_e \bar{\Phi}_0)$ 
cannot be a candidate, because it has $R=+1$ but it is 
not a U(3)$'$ singlet.]
Therefore, we can take superpotential 
$$
W= \lambda_1 [( P \bar{E} +E \bar{P}) \hat{\Theta}_P] +
\lambda_2 [(\Phi_e \bar{\Phi}_e + b\,  \Phi_0 \bar{\Phi}_0) \hat{\Theta}_P] ,
\eqno(4.3)
$$
so that we obtain
$$
k (P \bar{E} +E \bar{P}) = \Phi_e \bar{\Phi}_e + b\,  \Phi_0 \bar{\Phi}_0 , 
\eqno(4.4)
$$
i.e.
$$
2 k \cos \phi_i = x_i^4 + b\, x_i^2,
\eqno(4.5)
$$
where we have used the dimensionless expressions of $P$, $E$, $\Phi_0$ and 
$\Phi_e$, Eq.(2.9), Eq.(1.7) with $v_0=1$, and so on. 

Eliminating the coefficient $k$ in Eq.(4.5), we obtain 
two equations
$$
\frac{\cos \phi_1}{\cos \phi_3} = \frac{x_1^4+b\, x_1^2}{
 x_3^4+b\,  x_3^2} ,
 \eqno(4.6)
$$
$$
\frac{\cos \phi_2}{\cos \phi_3} = \frac{x_2^4+b\,  x_2^2}{
 x_3^4+b\,  x_3^2} .
 \eqno(4.7)
$$

In Sec.3, we have obtained numerical results $\phi_1= -41.815^\circ$ and 
$\phi_2= -15.128^\circ$ by putting $\phi_3$ as $\phi_3=0$. 
In order to avoid confusing, we use notation $\tilde{\phi}_i$ for these
numerical results of $\phi_i$.
Since we can choose any value of $\phi_0$ in $\phi_i \rightarrow \phi_i +\phi_0$, 
we define $\phi_i$ in Eq.(4.5) as
$$
\phi_1=\phi_0 + \tilde{\phi}_1, \ \ \phi_2=\phi_0 + \tilde{\phi}_2, \ \ 
\phi_3=\phi_0 .
\eqno(4.8)
$$
The equations (4.6) and (4.7) have two unknown parameters $\phi_0$ and $b$
under the input values $\tilde{\phi}_1$ and $\tilde{\phi}_2$.
So, we obtain
$$
\phi_0 = -45.903^\circ , \ \ \ \ b = -1.11586 ,
\eqno(4.9)
$$
which means
$$
\phi_1 = -87.718^\circ, \ \ \ \phi_2 = -61.031^\circ, \ \ \ 
\phi_3 = -45.903^\circ .
\eqno(4.10)
$$
 
Regrettably, since we need two input parameters $\phi_0$ and $b$
in order to predict the values $\tilde{\phi}_1$ and  $\tilde{\phi}_2$, 
the present model has no predictability for phase parameters
$(\phi_1, \phi_2, \phi_3)$. 
(If we use the fitting value $\tilde{\phi}_1 =-41.815^\circ$ as input value
in addition to the input value $b= -1.11586$, we can predict the value 
$\tilde{\phi}_2$ together with the value of $\phi_0$.)
However, note that the parameters $\phi_i$ are family number-dependent 
parameters, while the parameters $\phi_0$ and $b$ are
family number-independent parameters.
Therefore, the aim of the Yukawaon model that we understand 
mass spectra and mixings of all quarks and leptons only 
in terms of charged lepton mass spectrum and without 
using any other family number-dependent parameters
has been achieved in this scenario.

\vspace{5mm}

\noindent{\large\bf 5 \ Concluding remarks}

We have tried to describe quark and lepton mass matrices
by using only the observed values of charged lepton masses 
$(m_e, m_\mu, m_\tau)$ as input parameters with family 
number-dependent values. 
Namely, we have investigated  
whether we can describe all other observed mass spectra 
(quark and neutrino mass spectra) and the quark- and lepton-mixings (the CKM and the 
PMNS mixings) without using any other family number-dependent parameters. 
In conclusion, as seen in Sec.3, we have obtained reasonable 
results.
Our predicted values are listed in Table~3.

As seen in Sec.3, we have still used the phase matrix $P$ 
defined by Eq.(1.10)
in order to fit the observed CKM mixing parameters similarly to the  
past Yukawaon models. 
However, as seen in Sec.4, the most remarkable point of the present paper 
is that 
we have succeeded in describing the family number-dependent 
parameters $(\phi_1, \phi_2,\phi_3)$ by the family number-independent 
parameters $\phi_0$ and $b$.
Therefore, the main aim in the Yukawaon model have been 
achieved in the present work.  
However, regrettably, the mechanism proposed  has no 
predictability of the phase parameters $\phi_i$,
although it transforms unwelcome family number-dependent 
parameters into family number-independent parameters. 
The mechanism will be improved in a future version. 

The successful results in the present work suggests 
the following items:
(i) the flavor basis in which the charged lepton mass matrix
$M_e$ is diagonal is more fundamental basis in the flavor 
physics.
(ii) The parameters $(m_e, m_\mu, m_\tau)$ (i.e. $(x_1, x_2, x_3)$
defined by Eq.(1.7)) are fundamental parameters in quark and 
lepton physics. 
Note that the parameter values $(m_e, m_\mu, m_\tau)$ are
extremely hierarchical, while the parameter values 
$(x_1, x_2, x_3)$ are mildly hierarchical. 
Understanding of the values of $(x_1, x_2, x_3)$ will be left to 
our next task in future. 
Then, the relation $(m_e+m_\mu+m_\tau)/(\sqrt{m_e}+\sqrt{m_\mu}
+\sqrt{m_\tau})^2=2/3$ \cite{K-mass} may play an essential role
in investigating the origin of the parameter values $(x_1, x_2, x_3)$. 
For reference, we give a trial model on the charged lepton mass 
relation within the framework of the present Yukawaon model 
in Appendices A and B, although this is only a trial one.

In this model, there are four phase parameters $\alpha_\nu$, 
$\alpha_d$ and $(\tilde{\phi}_1, \tilde{\phi}_2)$. 
The parameters $\alpha_\nu$ and $\alpha_d$ play a role 
in giving mass ratios in the neutrino and down-quark sectors,
respectively.
The parameters which purely contribute to the CKM and PMNS 
mixing matrices as $CP$ violating phase parameters 
are only $(\tilde{\phi}_1, \tilde{\phi}_2)$. 
These parameters can commonly contribute to CKM and 
PMNS mixing matrix, so that those play an essential 
role in both the predicted values of $\delta_{CP}^q$ 
and $\delta_{CP}^\ell$. 
It is interesting that, in spite of different values between 
$\alpha_d$ and $\alpha_\nu$, the results of $CP$ violating 
parameters $\delta_{CP}^q$ and $\delta_{CP}^\ell$ take a similar
magnitude,  $\delta_{CP}^q \sim - \delta_{CP}^\ell \sim 70^\circ$.   

In conclusion, it seems to be certain that all of the observed 
hierarchical structures of quark and lepton masses and mixings 
are commonly originated from the hierarchical values of 
 $(m_e, m_\mu, m_\tau)$ which are described by the fundamental 
parameters $(x_1, x_2, x_3)$. 
Of course, the present Yukawaon model  has to be still improved 
with respect to  the $R$ charge assignments, number of flavons, 
number of adjustable parameters, $CP$ violating phase parameters, 
and so on. 
In addition to this, our next task is to investigate the origin of 
the parameters $(x_1, x_2, x_3)$.

\newpage

{\large\bf  Appendix A: Charged Lepton Mass Relation in the Yukawaon Model}

The charged lepton mass relation \cite{K-mass}
$$
K \equiv \frac{m_e+ m_\mu +m_\tau}{
(\sqrt{m_e} +\sqrt{m_\mu} + \sqrt{m_\tau})^2} = \frac{2}{3} ,
\eqno(A.1)
$$
is one of the main motives of the Yukawaon model in the earlier stage
 \cite{YK_MPL90}. 
The relation (A.1) can be understood from VEV of U(3) ${\bf 8}+{\bf 1}$ 
scalar, 
$\langle \hat{\Phi}_e \rangle = {\rm diag}(\sqrt{m_e}, 
\sqrt{m_\mu}, \sqrt{m\tau})$ as
$$
K= \frac{ {\rm Tr}[\hat{\Phi}_e \hat{\Phi}_e] }{
({\rm Tr}[\hat{\Phi}_e ])^2 } ,
\eqno(A.2)
$$
where we have omitted VEV notation ``$\langle$" and ``$\rangle$" 
for simplicity.
Also, hereafter, for simplicity, we denote ${\rm Tr} [A]$ as $[A]$.
However, in the present scenario of the Yukawaon model, there is 
no ${\bf 8}+{\bf 1}$ scalar $\hat{\Phi}_e$, but we have only 
${\bf 6}$ and ${\bf 6}^*$ scalars $\Phi_e$ and $\bar{\Phi}_e$. 
The purpose of the present paper is to understand mass ratios and 
mixings of quarks and leptons under the given parameters 
$(m_e, m_\mu, m_\tau)$, and it is not to investigate that the origin 
of the values $(m_e, m_\mu, m_\tau)$.

However, in this Appendix, let us try to understand the mass relation (A.1) 
according to an idea suggested in Ref.\cite{YK_MPL90}.
First, let us introduce ${\bf 8}+{\bf 1}$ scalar $\hat{\Phi}_e$. 
By using the following superpotential:
$$
W = \mu  [\hat{\Phi}_e \hat{\Theta}_e] +
\lambda_e  [(\Phi_e \bar{E} + E \bar{\Phi}_e)  \hat{\Theta}_e] ,
\eqno(A.3)
$$
we obtain a relation
$$
\hat{\Phi}_e =  \Phi_e \bar{E} + E \bar{\Phi}_e .
\eqno(A.4)
$$
Since $R(E) = \frac{1}{2}(1-\Delta)$ as seen in Eq.(2.19), 
$\hat{\Phi}_e$ has $R$ charge as 
$$
R(\hat{\Phi}_e) = 1 -\frac{1}{2} \Delta .
\eqno(A.5)
$$
Let us take $\Delta= +1$, so that we have
$$
R(\hat{\Phi}_e) = R(\Phi_e) = \frac{1}{2} .
\eqno(A.6)
$$
This choice (A.6) causes no problem because $\hat{\Phi}_e$ 
and $\Phi_e$ have different transformation under U(3)$\times$U(3)$'$. 
We will comment on the choice $R(E)=0$ later.

Since $R(\hat{\Phi}_e)=1/2$, we assume the following 
superpotential
$$
W= \frac{1}{\Lambda} \left( 
\lambda  [\hat{\Phi}_e \hat{\Phi}_e]^2 +
\lambda' [\hat{\Phi}_e] )^2 
[\hat{\Phi}_8 \hat{\Phi}_8] \right),
\eqno(A.7)  
$$
where $\hat{\Phi}_8$ is an octet part of the nonet $\hat{\Phi}_e$ 
defined by
$$
\hat{\Phi}_8 \equiv \hat{\Phi}_e - \frac{1}{3} 
[\hat{\Phi}_e]\ {\bf 1} .
\eqno(A.8)
$$

The first term in Eq.(A.7) is the conventional nonet-nonet term.
The second term is an (octet-octet)$\times$(singlet-singlet) interaction
term \cite{YK_MPL90} although the second term is still SU(3) invariant. 
In order to derivate the relation (A.1), the assumption of the second 
term is essential. 
By noticing that the second term can be expressed as
$$
[\hat{\Phi}_e \hat{\Phi}_e] [\hat{\Phi}_e]^2 
-\frac{1}{3} [\hat{\Phi}_e]^4,
\eqno(A.9)
$$
we obtain 
$$
\frac{\partial W}{\partial \hat{\Phi}_e} = \frac{1}{\Lambda} \left\{ 
2\left(2 \lambda [\hat{\Phi}_e \hat{\Phi}_e] + \lambda' [\hat{\Phi}_e]^2 
\right) \hat{\Phi}_e
+ 2 \lambda' \left( [\hat{\Phi}_e \hat{\Phi}_e ] 
-\frac{2}{3} [\hat{\Phi}_e]^2 \right) [\hat{\Phi}_e] {\bf 1} \right\}.
\eqno(A.10)
$$
The coefficients of $\hat{\Phi}_e$ and 
${\bf 1}$ must be zero in order to have a nontrivial solution of 
$\hat{\Phi}_e$ (non-zero and 
non-unit matrix form).
Thus,  we demand
$$
2 \lambda [\hat{\Phi}_e \hat{\Phi}_e] + \lambda' [\hat{\Phi}_e]^2=0 ,
\eqno(A.11)
$$
and
$$
[\hat{\Phi}_e \hat{\Phi}_e ] 
-\frac{2}{3} [\hat{\Phi}_e]^2  =0.
\eqno(A.12)
$$
Eq.(A.11) requires a special relation between $\lambda$ and
$\lambda'$. 
Note that the relation (A.12) is independent of the explicit 
value of $\lambda'$. 

Let us comment on the choice of  $\Delta=+1$. 
This choice means that  $R(E)=0$, so that  a U(3) nonet $(E\bar{E})$ takes $R(E\bar{E})=0$.
Therefore, the factor $E\bar{E}$ can be inserted into any terms 
with $R=2$ in the superpotential. 
However, since $\langle E \bar{E}\rangle ={\bf 1}$, this does not affect the obtained 
VEV relations practically. 
The choice $\Delta=+1$ also gives $R$ charges of $S_f$ as
$$
\left( R(S_\nu), R(S_d), R(S_e),  R(S_u) \right) 
= \left( \frac{1}{2}, 1, \frac{3}{2}, 2 \right). 
\eqno(A.13)
$$
It is interesting that the values $(1/2, 1, 3/2, 2)$ in (A.13) are the same 
as the values of  $\Phi_f$ as seen in Eq.(2.30), but the arrangements 
are different, i.e. $(e, u, \nu, d)$ for $R(\Phi_f)$, while $(\nu, d, e, u)$ 
for $R(S_f)$. 

\newpage

{\large\bf  Appendix B: Alternative Scenario for Charged Lepton Mass Relation}

In Appendix A, we have introduced the new flavon $\hat{\Phi}_e$ 
in addition to the flavons $\Phi_e$ and $\bar{\Phi}_e$.  
So far, we have adhered the idea that the Yukawaon VEV structures 
take a universal form $\hat{Y}_f = \Phi_f \bar{\Phi}_f + \xi_f  {\bf 1}$ 
($f=e, \nu, d, u$).
However, if we accept an idea that a structure of $\hat{Y}_e$ is 
exceptional, we can introduce $\hat{\Phi}_e$ without introducing 
$\Phi_e$ and $\bar{\Phi}_e$ as following
$$
\hat{Y}_e = \hat{\Phi}_e \hat{\Phi}_e , \ \ \ \ 
\hat{\Phi}_e = \Phi_0 \bar{\Phi}_0 .
\eqno(B.1)
$$
This expression (B.1) is rather simpler compared with the expression
$\hat{Y}_e = \Phi_e \bar{\Phi}_e$ with $\bar{P}_e \Phi_e \bar{P}_e =
\bar{\Phi}_0 S_e \bar{\Phi}_0$ given in Sec.2. 
Therefore, in this scenario, without $\Phi_e$ and $\bar{\Phi}_e$
[i.e. without Eqs.(A.3) - (A.6)], we can use (A.7),
so that we can obtain the charged lepton mass relation (A.1).

However, in this scenario, since we have $R$ charges 
$$
R(\hat{Y}_e)=+1, \ \ \ R(\hat{\Phi}_e)= \frac{1}{2} , \ \ \ 
R(\Phi_0)=\frac{1}{4} ,
\eqno(B.2)
$$
we cannot put the $\Phi_0 \bar{\Phi}_0$ term in Eq.(4.3).
In order to avoid this trouble, in a superpotential for 
$ P\bar{E}+E\bar{P}$, we a little change the scenario in Sec.4. 
We assume a mechanism similar to Eqs.(2.1) - (2.3): 
$$
W_P = \left[\left( \lambda_1 (P\bar{E} +E\bar{P}) +\lambda_2 \hat{\Phi}_e
\hat{\Phi}_e \right) \hat{\Theta}_P \right] +
\left[ \lambda'_1(P\bar{E} +E\bar{P}) +\lambda'_2 \hat{\Phi}_e \hat{\Phi}_e 
\right] [ \hat{\Theta}_P], 
\eqno(B.3)
$$
so that we obtain 
$$
k (P\bar{E}+E\bar{P}) = \hat{\Phi}_e \hat{\Phi}_e + \xi_P {\bf 1} ,
\eqno(B.4)
$$
i.e.
$$
2 k \cos \phi_i = x_i^4 + \xi_P ,
\eqno(B.5)
$$
instead of Eqs.(4.3) and(4.4), respectively. 
From Eqs.(4.5) and (4.6) with $\xi_P$ instead of $b$ terms, 
we obtain numerical solution 
$$
\phi_0=29.222^\circ , \ \ \ \ \xi_P=-5.9619 ,
\eqno(B.6)
$$
so that 
$$
\phi_1= -12.623^\circ, \ \ \ \phi_2=14.069^\circ, \ \ \ \phi_3 = 29.222^\circ.
\eqno(B.7)
$$ 

Maybe, other scenarios are also possible.
The purpose in this paper is not to propose a scenario which 
derives the relation (A.1) 
but to demonstrate a possibility that the family number-dependent 
parameters $(\phi_1, \phi_2, \phi_3)$ can, 
in principle, be described by a family number-independent parameter. 
More reasonable scenario will be given in future.

\vspace{2mm}
\newpage
%


\begin{thebibliography}{99} 
%
\bibitem{f-symmetry} 
K.~Akama and H.~Terazawa, INS-Report-257 (1976) (INS, University of
Tokyo);  H.~Terazawa, Y.~Chikashige, and K.~Akama, Phys.Rev. {\bf D15}, 
 480 (1977); T.~Maehara and T.~Yanagida, Prog.~Theor.~Phys. {\bf 60}, 
822 (1978).
%
\bibitem{YK_09}
Y.~Koide, Phys~Rev. {\bf D 79}, 033009 (2009); Phys.~Lett. {\bf B 680}, 76 (2009).
%
%
\bibitem{flavon}  C.~D.~Froggatt and H.~B.~Nelsen, Nucl.~Phys. 
{\bf B 147}, 277 (1979).
For recent works, for instance, see R.~N.~Mohapatra, 
AIP Conf.~Proc. {\bf 1467}, 7 (2012);
A.~J.~Buras {\it et al.}, JHEP {\bf 1203} (2012) 088. 
%
\bibitem{seesaw}
P.~Minkowski, Phys.~Lett. {\bf B 67}, 421 (1977); 
M.~Gell-Mann, P.~Ramond, and R.~Slansky, Proceedings of the
Supergravity Stony Brook Workshop, New York, 1979,
edited by P.~Van Nieuwenhuizen and D.~Freedman
(North-Holland, Amsterdam, 1979); 
T.~Yanagida, Proceedings of the Workshop on Unified Theories and
Baryon Number in the Universe, Tsukuba, Japan 1979,
edited by A. Sawada and A. Sugamoto [KEK Report No.~79-18, Tsukuba]; 
R.~Mohapatra and G.~Senjanovic, Phys.~Rev.~Lett. {\bf 44}, 912 (1980).
%
%
\bibitem{CKM} N.~Cabibbo, Phys.~Rev.~Lett. {\bf 10}, 531 (1963); 
M.~Kobayashi and T.~Maskawa, Prog.~Theor.~Phys. {\bf 49}, 
 652 (1973).
%
\bibitem{Yukawaon_1}
H.~Nishiura and Y.~Koide, Phys.~Rev. {\bf D 83}, 035010 (2011); 
Euro.~Phys.~J.  {\bf C 72}, 1933 (2012); 
Phys.~Lett. {\bf B 712}, 396 (2012).
%
\bibitem{tribi} 
P.~F.~Harrison and W.~G.~Scott, Phys.~Lett. {\bf B 535}, 163 (2002); 
Phys.~Lett. {\bf B 557}, 76 (2003); 
Z.-z.~Xing, Phys.~Lett. {\bf B 533}, 85 (2002); 
E.~Ma, Phys.~Rev.~Lett. 90, 221802 (2003); 
C.~I.~Low and R.~R.~Volkas, Phys.~Rev. {\bf D 68}, 033007 (2003). 

%
\bibitem{Yukawaon_2}
Y.~Koide and H.~Nishiura, Euro.~Phys.~J.  {\bf C 73}, 2277 (2013); 
 JHEP  {\bf 04}, 166 (2013); 
 Phys.~Rev. {\bf D 88}, 116004 (2013).
%
\bibitem{theta13}
K.~Abe {\it et al.} (T2K collaboration),
Phys.~Rev.~Lett. {\bf 107}, 041801 (2011);
%
MINOS collaboration, P. Adamson {\it et. al.}, 
 Phys.~Rev.~Lett. {\bf 107}, 181802 (2011);
%
Y. Abe {\it et al.}  (DOUBLE-CHOOZ Collaboration), 
Phys.~Rev.~Lett. {\bf 108}, 131801  (2012);
%
F.~P.~An, {\it et al.} (Daya-Bay collaboration),  
 Phys.~Rev.~Lett. {\bf 108}, 171803 (2012);
%
J.~K.~Ahn, {\it et al.} (RENO collaboration),  
 Phys.~Rev.~Lett. {\bf 108}, 191802 (2012).
%
%
\bibitem{Yukawaon_3} 
Y.~Koide and H.~Nishiura, Phys.~Rev. {\bf D 90}, 016009 (2014); 
 Phys.~Rev. {\bf D 90}, 117903 (2014). 
%
\bibitem{K-F_ZPC96} 
Y.~Koide and H.~Fusaoka, Z.~Phys. {\bf C 71}, 459 (1996).
%
%
\bibitem{PMNS} 
B.~Pontecorvo, Zh.~Eksp.~Teor.~Fiz. {\bf 33}, 
 549 (1957) and {\bf 34}, 247 (1957); 
Z.~Maki, M.~Nakagawa, and S.~Sakata, Prog.~Theor.~Phys. {\bf 28},  
 870 (1962).
%
%
\bibitem{q-mass} Z.-z.~Xing, H.~Zhang, and S.~Zhou, 
{Phys.~Rev.} {\bf D 77}, 113016 (2008).
%
\bibitem{q-mass2}
H.~Fusaoka and Y.~Koide, {Phys. Rev.} 
{\bf D 57}, 3986 (1998).
%
%
\bibitem{PDG14} 
  K.~A.~Olive {\it et al.} (Particle Data Group),
 Chinese Phys. C, {\bf 38}, 09001 (2014).
%
%
\bibitem{UTfit} 
UTfit Collaboration, Fit results: Summer 2014 at \\
http://www.utfit.org/UTfit/ResultsSummer2014PostMoriondSM
%
\bibitem{Doi1981} M.~Doi, T.~Kotani, H.~Nishiura, K.~Okuda, and E.~Takasugi,
 Phys.~Lett. {\bf B 103}, 219 (1981) and {\bf B 113}, 513 (1982).
%
%
\bibitem{K-mass} 
  Y.~Koide,
  Lett.\ Nuovo Cim.\  {\bf 34},  201 (1982);
%
  Phys.\ Lett.\  B {\bf 120},  161 (1983);
%
  Phys.\ Rev.\  D {\bf 28},  252(1983).
%
\bibitem{YK_MPL90}
 Y.~Koide, Mod.~Phys.~Lett. {\bf A5}, 2319  (1990).
 Also see, Y.~Koide, Phys~Rev. {\bf D 79}, 033009 (2009).  
%


\end{thebibliography}
\end{document}